%% file: main.tex
\documentclass{jpp}
\usepackage{graphicx}
\usepackage{xcolor}
\usepackage[utf8]{inputenc}
\usepackage[T1]{fontenc}
\usepackage{amsmath}
\usepackage{hyperref}
\usepackage[capitalize]{cleveref}

\usepackage[normalem]{ulem}
\newcommand\redout{\bgroup\markoverwith
{{\rule[.5ex]{2pt}{0.4pt}}}\ULon}

\shorttitle{A single-field-period quasi-isodynamic stellarator}
\shortauthor{R. Jorge {et al.}}

\title{A single-field-period quasi-isodynamic stellarator}

\author{R. Jorge\aff{1},
G. G. Plunk\aff{1},
M. Drevlak\aff{1},
M. Landreman\aff{2},
J.-F. Lobsien\aff{1},
K. Camacho Mata\aff{1}
\and P. Helander\aff{1}
}

\affiliation{\aff{1}Max-Planck-Institut für Plasmaphysik, D-17491 Greifswald, Germany
\aff{2}Institute for Research in Electronics and Applied Physics, University of Maryland, College Park, MD 20742, USA}

\begin{document}

\maketitle

\begin{abstract}
A single-field-period quasi-isodynamic stellarator configuration is presented. This configuration, which resembles a twisted strip, is obtained by the method of direct construction, that is, it is found via an expansion in the distance from the magnetic axis.  Its discovery, however, relied on an additional step involving numerical optimization, performed within the space of near-axis configurations defined by a set of adjustable magnetic-field parameters. This optimization, completed in 30 seconds on a single cpu core using the SIMSOPT code, yields a solution with excellent confinement, as measured by the conventional figure of merit for neoclassical transport, effective ripple, at a modest aspect ratio of eight. The optimization parameters that led to this configuration are described, its confinement properties are assessed, and a set of magnetic-field coils is found. The resulting transport at low collisionality is much smaller than that of W7-X, and the device needs significantly fewer coils thanks to the reduced number of field periods.
\end{abstract}

\section{Introduction}
\label{sec:intro}

In 1951, Lyman Spitzer proposed the stellarator concept, where the magnetic field lines are twisted by deforming a torus to break axisymmetry \citep{Spitzer1958}.
Spitzer's idea involved the shape of a figure eight, where he was able to calculate its corresponding rotational transform, and was realized experimentally with the Model A, B and C stellarators \citep{Stix1998}.
During the next several decades, the stellarator concept saw important developments in the calculation of MHD equilibria with good nested flux surfaces, high-$\beta$ stability properties, coil optimization, reduced neoclassical transport, and other improvements \citep{Gates2017}.
More recently, the figure-eight configuration proposed by Spitzer has continued to inspire advancements in the design of stellarators and linked mirrors \citep{Feng2021}.

%
There are currently three types of optimized stellarator configurations that appear to have the potential to become future fusion reactors: (1) quasi-isodynamic (QI) \citep{Gori1994}, (2) quasi-axisymmetric (QA) and (3) quasi-helically symmetric (QH) ones \citep{Boozer2015a}.
In these configurations, {termed omnigenous}, the second (or longitudinal) adiabatic invariant, $J$, is a function of the toroidal magnetic flux only, $J = J(\psi)$, which can be shown to lead to confined guiding center orbits \citep{Helander2014}.
The difference between the three cases lies mainly in the way that the contours of constant magnetic field strength $|\mathbf B|$ close when plotted on magnetic flux surfaces: in QI they close poloidally, in QA toroidally, and in QH helically.
Additionally, while QA and QH belong to the class of quasisymmetric stellarators, where all contours of $|\mathbf B|$ are straight in Boozer and Hamada coordinates \citep{d1991flux}, in QI only the contours of the maxima of $|\mathbf B|$ are required to be straight.
All three types of configurations can be ``directly constructed'' using the near-axis expansion, based on an approach using Boozer coordinates \citep{Garren1991,Landreman2019a, Plunk2019}, while only quasisymmetric stellarators have been obtained using an approach based on Mercier coordinates \citep{Jorge2020,Jorge2020b}.

The largest stellarator in operation,  Wendelstein 7-X \citep{Grieger1992b}, is of the QI type, though only very approximately {given the fact that the magnetic field strength at local maxima and minima varies substantially over the flux surface, there is substantial neoclassical transport, and losses of fast ions and the bootstrap current is non-negligible}.

This type of configuration is relatively insensitive to the plasma pressure since the Shafranov shift is small \citep{Gori1994,Nuhrenberg1995,Cary1997,Nuhrenberg2010a} and the bootstrap current vanishes identically at low collisionality \citep{Helander2009,Helander2011,Landreman2012}.
Unlike quasisymmetric configurations, quasi-isodynamic ones may thus be operated with essentially no net toroidal current {experimentally, even with finite plasma pressure}.

In this work, we construct a QI configuration using a near-axis expansion framework based on Boozer coordinates, which reduces the computational effort considerably compared with other approaches (the advantages and limitations of the near-axis expansion framework are detailed in the next section).
Unlike nearly all previous stellarator designs, our configuration only has a single field period and a racetrack shape that resembles Spitzer's original idea.
In contrast to the latter, however, it has carefully shaped flux surfaces in order to satisfy the stringent requirements of quasi-isodynamicity. 

As shown recently \citep{Landreman2019b, Landreman2020a, Jorge2020b, Landreman2021a}, the near-axis expansion can be used not only as a tool to find configurations with enhanced particle confinement, but it also has the potential to find configurations that are Mercier stable, if the expansion is carried to second order.
Our approach, {based on a first-order near axis expansion}, relies on a careful choice of these parameters such that good flux surfaces are found even for low to medium aspect ratios, i.e., for $R/a$ between 6 and 10, which are within the realm of the near-axis expansion framework.
In the past, QI stellarators have usually had aspect ratios of at least 10 {(as for example the W7-X A configuration of \citet{Geiger2015} at $\beta=0$)}.

The quasi-isodynamic configuration found here is obtained by varying the input parameters for the near-axis expansion, namely the magnetic axis and the zeroth and first order magnetic fields, and optimizing the resulting output such that a given objective function is minimized.
This optimization is performed using the SIMSOPT code \citep{Landreman2021,Landreman2021b}{, which is able to perform gradient-based optimizations with MPI parallelization of a finite-difference method and can easily be generalized to a VMEC-based optimization approach that uses near-axis configurations as initial conditions.}
After a suitable set of parameters is found, a finite-aspect-ratio configuration is constructed and its properties are assessed using several numerical tools commonly used in stellarator optimization studies: the Variational Moments Equilibrium Code (VMEC) \citep{Hirshman1986}, the BOOZ\_XFORM code \citep{Sanchez2000a}, the NEO code \citep{Nemov1999}, the Stepped Pressure Equilibrium Code (SPEC) \citep{Hudson2012,Qu2020}, the coil optimization suite ONSET \citep{ONSET} and  the SIMPLE code \citep{Albert2020}.
Their main functions and results will be described in the following sections.
The method used to construct near-axis QI fields is that of \citet{Plunk2019}, although we choose different forms of some of the free functions that parameterize the solution space. The most notable example is a function that controls the deviation from omnigenity near maxima of the magnetic field strength (see Eqn.~\ref{eq:alphanearaxis}); this change, which improves the smoothness of solutions, will be explained in more detail in an upcoming publication \citep{Camacho2022}.

This paper is organized as follows.
In \cref{sec:nearaxis} the near-axis expansion formalism is outlined, in particular its application to QI fields and the corresponding numerical implementation.
In \cref{sec:figures_of_meric} the physics-based figures of merit are described and their analytical expressions in the near-axis expansion formalism are shown.
The resulting optimized configuration is shown in \cref{sec:optimized_configurations}, and our conclusions follow in the final section. 

\section{The Near-Axis Expansion}
\label{sec:nearaxis}

The near-axis expansion solves the equilibrium MHD equations by performing an expansion in the inverse aspect ratio $\epsilon$,
%
%
\begin{equation}
    \epsilon=\frac{a}{R}\ll 1,
\end{equation}
where $a$ and $R$ are measures of the minor and major radius of the device, respectively.
{We note that, although the construction is based on an expansion on $\epsilon$, it is able to describe the core region of any configuration, including those with low aspect ratio.}
We employ the near-axis expansion using Boozer coordinates $(\psi,\theta,\varphi)$ \citep{Boozer1981}, with $\psi$ the toroidal magnetic flux divided by $2\pi$, $\theta$ the poloidal angle and $\varphi$ the toroidal angle, writing the magnetic field vector as
\begin{align}
    \mathbf B &= \nabla \psi \times \nabla \theta + \iota(\psi) \nabla \varphi \times \nabla \psi\\
    &= \beta(\psi,\theta,\varphi) \nabla\psi + I(\psi) \nabla \theta + G(\psi) \nabla \varphi,
\label{eq:boozerB1}
\end{align}
where $\iota=\iota(\psi)$ is the rotational transform, $I(\psi)$ is $\mu_0/(2\pi)$ times the toroidal current enclosed by the flux surface, $G(\psi)$ is $\mu_0/(2\pi)$ times the poloidal current outside the flux surface and $\beta$ is related to the Pfirsch-Schl\" uter current.

In the near-axis expansion method for quasi-isodynamic configurations, the position vector $\mathbf{r}$ is written as
\begin{align}
\label{eq:positionVector}
\mathbf{r}(r,\theta,\varphi) = \mathbf{r}_0(\varphi)
+X(r,\theta,\varphi) \mathbf{n}^s(\varphi)
+Y(r,\theta,\varphi) \mathbf{b}^s(\varphi)
+Z(r,\theta,\varphi) \mathbf{t}^s(\varphi),
\end{align}
where $r=\sqrt{2 \psi/\overline B}$ is a radial-like variable, $\overline{B}$ is a constant reference field strength, and $\mathbf{r}_0(\varphi)$ is the magnetic axis curve parametrized using $\varphi$.
We employ a modified Frenet-Serret frame \citep{Carroll2013} where the curvature is replaced by the signed curvature $\kappa^s=s \kappa$ where $s(\varphi)$ takes values of $\pm 1$ and switches at locations of zero {axis} curvature {(a characteristic of QI configurations)}.
The normal and binormal vectors are also multiplied by $s$.
Note that the Frenet-Serret formulas are invariant under such substitution.
We therefore write the signed Frenet-Serret frame as $(\mathbf{t},\mathbf{n}^s=s \mathbf n,\mathbf{b}^s=s \mathbf b)$, where $(\mathbf{t},\mathbf n, \mathbf b)$ are the tangent, normal and binormal unit vectors of the Frenet-Serret frame of the magnetic axis.
In the following, the arc length along the axis is denoted by $\ell$ with $0\le \ell < L$, the axis curvature by $\kappa(\varphi)$ and the axis torsion by $\tau(\varphi)$ {with the sign convention of \citet{Landreman2019}.}

We expand the magnetic field and the position vector only up to first order in $r$ as {\citep{Garren1991}}
\begin{equation}
    B=B_0[1+r d \cos(\theta-\alpha)],
\label{eq:Bnearaxis}
\end{equation}
{where $B_0=B_0(\varphi)$ is the magnetic field on-axis, $d=d(\varphi)$ is a free function describing the first order magnetic field strength, while $\alpha=\alpha(\varphi)$ is an angle-like variable also describing the first order magnetic field strength}.
This yields \citep{Landreman2019b}
\begin{align}
    X = \frac{r d}{\kappa^s} \cos(\theta-\alpha),~
    Y = \frac{r \kappa^s }{d}\frac{\overline B}{B_0}\left[\sin(\theta-\alpha)+\sigma \cos(\theta-\alpha)\right],~
    Z = 0,
\end{align}
where $G= G_0 = L/\int_0^{2\pi}(d\varphi/B_0)$ and $\sigma$ is a solution of
\begin{equation}
    \frac{d \sigma}{d\varphi}+\gamma\left(1+\sigma^2+\frac{B_0^2}{\overline B^2}\frac{d^4}{(\kappa^s)^4}\right)-2\left(\frac{I_2}{\overline B}-\tau\right)\frac{G_0 d^2}{\overline B (\kappa^s)^2}=0,
\label{eq:sigma_omng}
\end{equation}
with {$I=r^2 I_2$}
and $\gamma=\iota-\alpha'(\varphi)$.
As the plasma pressure only appears at second order in the expansion, the configurations considered here are effectively {force-free} configurations.
{Incidentally, the fact that the axis curvature should vanish at points where $B_0'(\varphi)=0$ stems from the fact that QI fields need to have $d=0$ at all local extrema combined with $Y$ being proportional to $r \kappa/d$ \citep{Plunk2019}.}

As shown by \citet{Cary1997}, QI fields are necessarily non-analytic.
Furthermore, for the near-axis expansion case, the omnigenity condition leads to the relation $\alpha(2\pi)-\alpha(0)=2 \pi \iota$, while periodicity of the magnetic field in \cref{eq:Bnearaxis} requires $\alpha(2\pi)-\alpha(0)=2 \pi N$ \citep{Plunk2019}.
To alleviate this conflict between omnigenity and periodicity, we choose the function $\alpha$ such that omnigenity is violated in a controlled way, by writing it as
\begin{equation}
    \alpha = \iota\left(\varphi-\frac{\pi}{N_{fp}}\right)\left[1+\frac{\pi}{\iota}\left(\frac{N-\iota/N_{fp}}{(\pi/N_{fp})^{2k+1}}\right)\left(\varphi-\frac{\pi}{N_{fp}}\right)^{2k}\right]+\pi\left(2N+\frac{1}{2}\right),
\label{eq:alphanearaxis}
\end{equation}
where $N_{fp}$ is the number of field periods of the device.
The integer $k$ in \cref{eq:alphanearaxis} effectively controls the spatial distribution along $\varphi$ of the deviation from omnigenity.
{For a more detailed discussion and a comparison with other forms of $\alpha$ including the exact omnigenous form} see \citet{Camacho2022}.

We consider magnetic fields with a single minimum along the magnetic axis of the form
\begin{equation}
    B_0 = B_{00}+B_{01} \cos(N_{fp}\varphi),
\end{equation}
and parametrize the axis curve $\mathbf r_0$ as
\begin{equation}
    \mathbf r_0(\phi) = \sum_l R_l \cos(N_{fp} l \phi) \mathbf e_R + \sum_l Z_l \sin(N_{fp} l \phi) \mathbf e_Z,
\end{equation}
where $\phi$ is the standard cylindrical toroidal angle, $\mathbf e_R$ and $\mathbf e_Z$ are the standard cylindrical coordinate unit vectors and stellarator-symmetry is assumed.
The form for $d$ chosen here is similar to the approach taken by \citet{Plunk2019}, with an added term proportional to the curvature of the magnetic axis,
\begin{equation}
    d = d_\kappa \kappa + \sum_l \overline d_l \sin(N_{fp} l \varphi),
\label{eq:dnearaxis}
\end{equation}
where $d_\kappa$ and $\overline d_i$ are constants.

The equation for $\sigma$, \cref{eq:sigma_omng}, is solved using Newton iteration with a pseudo-spectral collocation discretization together with the constraint $\sigma(\phi=0)=\sigma_0$.
In the following, the parameter $\sigma_0$ will be set to $0$ in order to enforce stellarator symmetry.
If $\phi=0$ is not one of the grid points, this condition is imposed by interpolating $\sigma$ using pseudospectral interpolation.
A uniform grid of $N_\phi$ points is used, with $\phi_j=(j+j_{shift}-1)2\pi/(N_\phi N_{fp})$ for $j=1 \ldots N_\phi$ and $j_{shift}=0$ and $1/3$ in the quasisymmetric and quasi-isodynamic cases, respectively.
{The value of $j_{shift}=1/3$ is used to avoid points along the axis where the curvature reaches zero but different values of $j_{shift}$ could be used to achieve the same effect.}
The discrete unknowns include the values of $\sigma$ on the $\phi$ grid and $\iota_0$.
As a boundary condition, we impose periodicity in $\sigma$, with $\sigma(0)=\sigma(2\pi)=\sigma_0$, yielding a single value of $\iota_0$ and the function $\sigma(\phi)$ as solution.
Finally, a conversion to cylindrical coordinates is performed in order to create VMEC and SPEC input files and for visualization purposes using the non-linear method described in \citet{Landreman2019a} and by choosing a particular value for the radius $r$.

\section{Optimization Method}
\label{sec:figures_of_meric}

The optimization is performed with the SIMSOPT code, using the trust region reflective algorithm for nonlinear least squares problems from the \textit{scipy} package {using the Python programming language}.
%
%
The input parameters for the optimization are the axis shape coefficients $(R_i,Z_i)$ except $R_0=1$, which is fixed, the magnetic field on-axis, where we fix $B_{00}=1$ and vary $B_{01}$, and the scalars $d_\kappa$ and $\overline d_i$ present in the first-order magnetic-field function $d$ in \cref{eq:dnearaxis}.
The additional parameters $N_{fp}$, the number of field periods, and the exponent $k$ in \cref{eq:alphanearaxis} were varied manually.
As it was seen that values of $k=3$ and $N_{fp}=1$ yielded configurations with consistently lower elongation and neoclassical transport (as measured by $\epsilon_{\text{eff}}$; see below), these parameters were then held fixed during the construction of the configuration presented here.

The optimization is performed in a series of steps.
As an initial condition, we employ $R_0=1$, {$R_2=-0.2$}, $Z_2=0.35$,  $(R_1,Z_1)=0.0$, $B_{01}=0.16$ (a value used later as reference), $d_\kappa=0.5$, $\overline d_0=0$ and $\overline d_1=0.01$.
{The values of $R_0, R_2$ and $Z_2$ are the ones of the configuration in \citet{Plunk2019}}.
Then, SIMSOPT is called inside a loop over the number of axis coefficients, starting at only two coefficients up to {12} sequentially, that is, there are a total of {12} free axis coefficients, $R_2, R_4, R_6, R_8, R_{10}, R_{12}, Z_2, Z_4, Z_6, Z_8, Z_{10}, Z_{12}$, and therefore there are 6 steps, where first $R_2, Z_2$ is allowed to vary, then $R_4, Z_4$, then $R_6, Z_6$ {and so forth}.
{We note that, as more axis coefficients are introduced, the previous axis coefficients are still varied within the optimization.}
By choosing even mode numbers only, we obtain a 2-field period axis shape with two points of zero curvature at the points of the extrema of $|\mathbf B|$ ($\varphi=0$ and $\varphi=\pi$), as required by the omnigenity condition \citep{Plunk2019}.
The grid resolution is $N_\phi=131$ and is increased by 20 each time the number of axis coefficients increases.
Each iteration inside the loop is run until the change of the cost function is smaller than $10^{-4}$, which usually takes between 10 and 40 steps to be achieved.
The optimization process takes less than thirty seconds using a single CPU core.

The objective function has the following form
\begin{align}
    f_{QI}&=\frac{w_{L_{\nabla B}}}{L_{\nabla \mathbf B}^2}+w_{\mathrm{E}} |\mathrm{E}|^2+w_{R_0}[\text{min}(R_{axis})-R_{\text{min}}]^2+w_{Z_0}[\text{max}(Z_{axis})-Z_{\text{max}}]^2\nonumber\\
    &+w_d |d|^2+w_{\overline d_s}\sum_i |\overline d_i|^2+w_{\overline B_{01}}|\overline B_{01}-\overline B^{\text{ref}}_{01}|^2+w_{d'(0)}|d'(0)|^2+w_{\alpha}|\alpha-\alpha_0|^2.
\label{eq:fqi}
\end{align}
In \cref{eq:fqi}, $L_{\nabla \mathbf B}$ is the scale length associated with the Frobenius norm $|\nabla \mathbf B|$ of the $\nabla \mathbf B$ tensor (Eq. (3.11) in \citet{Landreman2021a}), given by $L_{\nabla \mathbf B} = B_0 \sqrt{2 / |\nabla \mathbf B|}$, $\text{min}(R_{axis})$ and $\text{max}(Z_{axis})$ are the value of the minimum cylindrical radial and maximum vertical coordinates, respectively, of the axis curve $\mathbf r_0$, $|d|$ and $|\mathrm{E}|$ are the $L2$ norm of the discretized functions $d$ and the elongation $\mathrm{E}=a/b$ associated with the first order magnetic field in \cref{eq:Bnearaxis} where $a$ and $b$ are the semi-major and semi-minor axis of the elliptical flux surface cross-section, respectively, and $\alpha_0=\iota(\varphi-\pi)$ is an exactly omnigenous version of the function $\alpha$ in \cref{eq:alphanearaxis}.
{The terms in \cref{eq:fqi} have three main goals: 1) select configurations with small deviations from $QI$ (min of $\alpha-\alpha_0$, $\overline d$, $d$ and $d'(0)$); 2) select axis shapes with small elongation and aspect ratio (reduction of $1/L_{\nabla \mathbf B}$, min($R_{axis}$), max($Z_{axis}$) and $E$; 3) penalize high mirror ratios (reduction of $\overline B_{01}-\overline B_{01}^{ref}$).}
{For a more in-depth assessment of the relation between axis shapes and the resulting elongation and the difficulty of obtaining solutions with low elongation in the near-axis expansion framework, we refer the reader to \cite{Camacho2022}.}
As parameters in \cref{eq:fqi} we choose $R_{\text{min}}=Z_{\text{max}}=0.4$ and $\overline B_{01}^{\text{ref}} = 0.16$.
The weights used in the optimization process are the following:
$w_{L_{\nabla B}}=0.03,w_e=0.4/N_{\phi},w_{R_0}=w_{Z_0}=30,w_d=20/N_{\phi},w_{\overline d_s}=100,w_{\overline B_{01}}=200,w_{d'(0)}=2,w_{\alpha}=60$.
{Such values were found by first scaling the weights such that every term in the objective function has an order of magnitude of unity when a reasonable equilibrium is found and then they are fine tuned to ensure the three main goals described before.}

\section{Results}
\label{sec:optimized_configurations}

The optimization procedure resulted in the following parameters
%
\begin{align}
    R_0 (\phi)~[\text{m}] &= 1 - 0.40566229 \cos(2\phi)+0.07747378 \cos (4 \phi)-0.00780386 \cos(6 \phi),\\
    Z_0 (\phi)~[\text{m}] &=-0.24769666 \sin(2\phi)+0.06767352 \sin(4 \phi)-0.00698062\sin(6\phi)\nonumber\\
    &-6.8162\times 10^{-4}\sin(8\phi)-1.451\times 10^{-5}\sin(10\phi)-2.83\times 10^{-6}\sin(12\phi),\\
    B_0(\phi)~[\text{T}] &= 1+0.16915531 \cos(\varphi),\\
    d (\phi) &= 0.00356311 \sin(2\varphi)+0.00020159\sin(4\varphi)-0.00121786\sin(6\varphi)\nonumber\\
    &-0.00011629\sin(8\varphi)-8.3\times 10^{-7}\sin(10\varphi)+3.20\times 10^{-6}\sin(12\varphi)\nonumber\\
    &+0.51837838\kappa.
\end{align}
This configuration has a rotational transform on-axis of $\iota_0 = 0.680$, a maximum elongation of $\mathrm{E}=5.100$, a derivative of $d$ on axis of $d'(0)=0.094$ and a total value of the objective function of $f_{QI}=32.618$.
The coefficients $R_8, R_{10}$ and $R_{12}$ are smaller than $10^{-8}$ and were therefore set to zero as they have a negligible contribution to the properties of the equilibrium.

Data for the magnetic configurations are available at \citet{Jorge2022}.

\begin{figure}
    \centering
    \includegraphics[width=.45\textwidth]{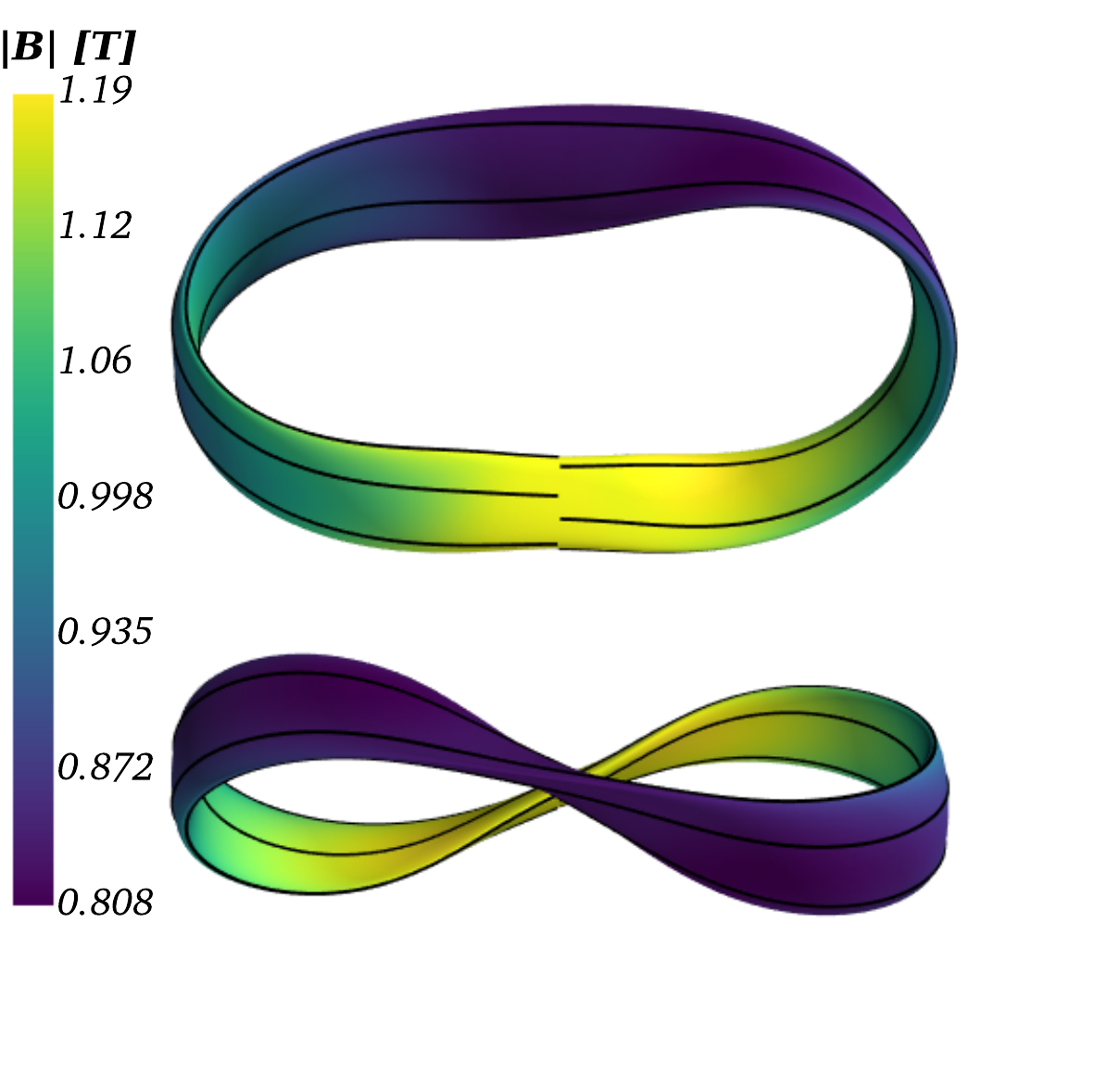}
    \includegraphics[width=.54\textwidth]{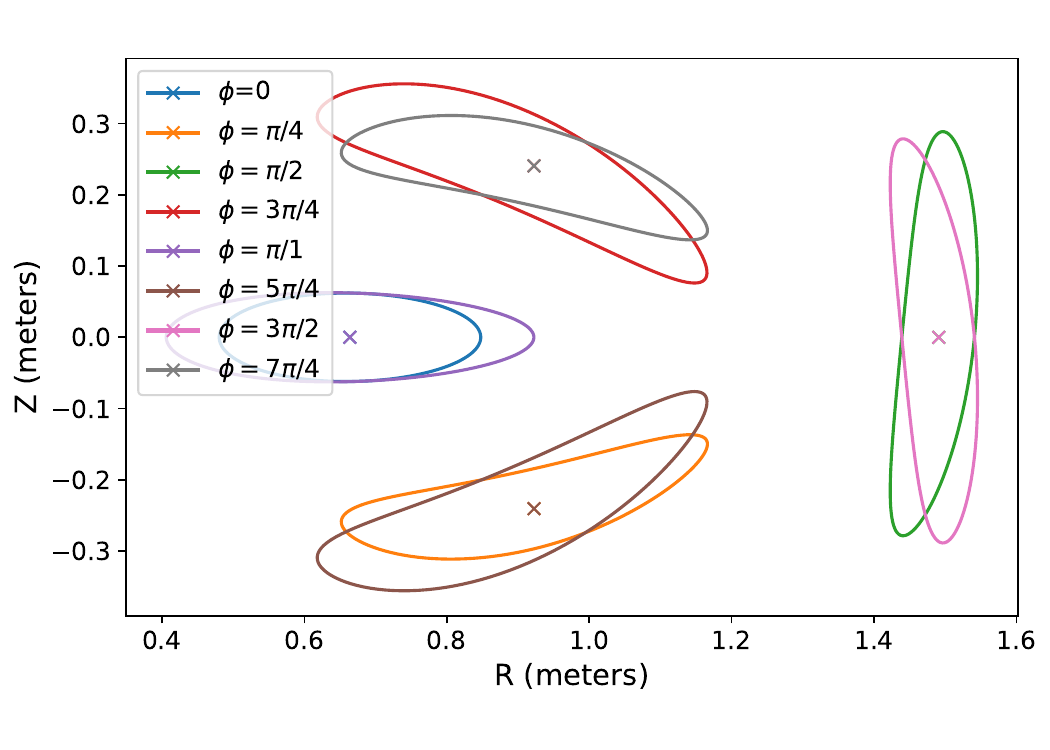}
    \caption{Shape of the idealized constructed configuration from the near-axis expansion. Left: Bird's eye and side view of the boundary shape in 3D with an aspect ratio of 8. Right: Cross-sections of the configuration at 8 values of toroidal angle $\phi$.}
    \label{fig:3d_config}
\end{figure}

\begin{figure}
    \centering
    \includegraphics[width=.49\textwidth]{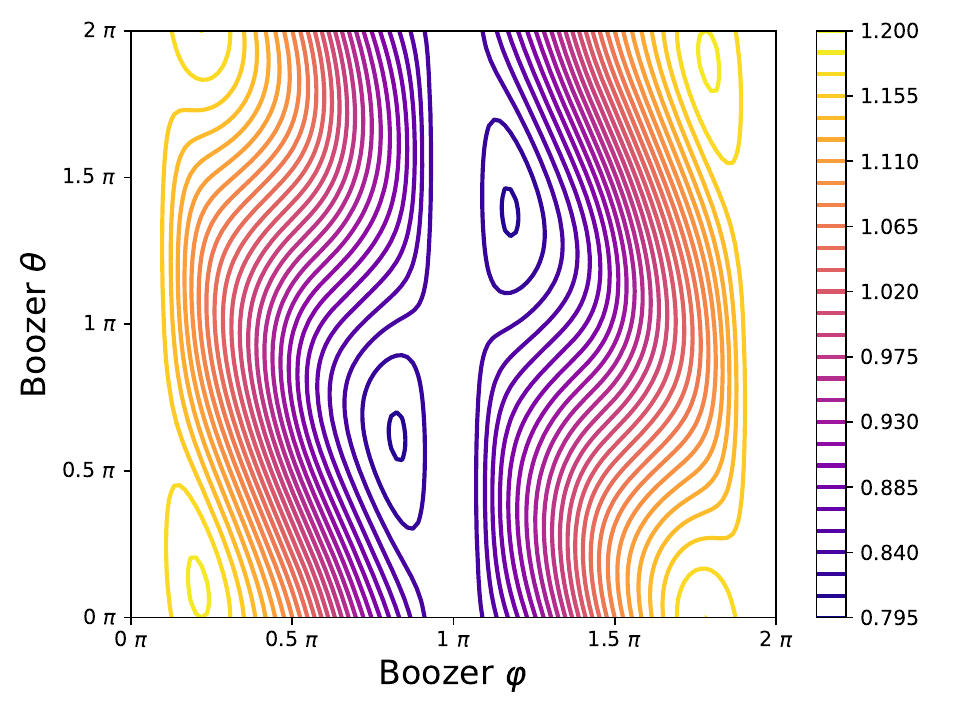}
    \includegraphics[trim=60 10 80 160,clip,width=.49\textwidth]{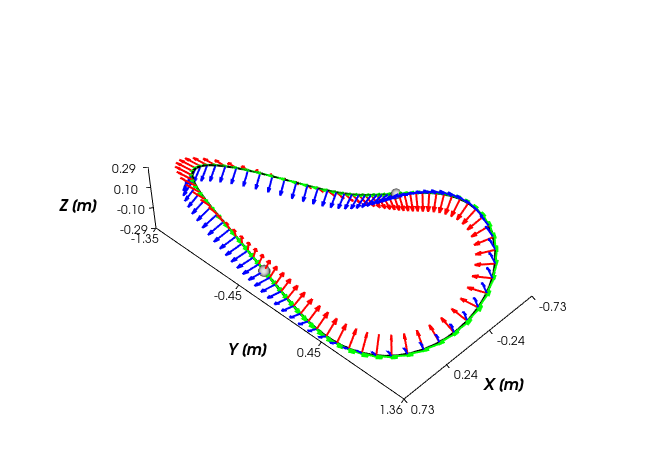}
    \caption{Elements of the near-axis construction. Left: Magnetic field on the boundary using \cref{eq:Bnearaxis}. Right: Shape of the magnetic axis with white spheres in the locations of zero curvature. In red we depict the signed normal vector, in blue the signed binormal vector and in green the tangent vector.}
    \label{fig:elements}
\end{figure}

To create a plasma boundary, we chose the radial near-axis coordinate as $r=1/9$, which leads to a calculated minor and major radius of Aminor\_p = 0.125 {(corresponding VMEC output parameter)} and Rmajor\_p = 0.993 {(corresponding VMEC output parameter)}, respectively, leading to an aspect ratio of $\epsilon$ = 7.944.
{The resulting} boundary can be seen in \cref{fig:3d_config}, while the magnetic field and the magnetic axis can be seen in \cref{fig:elements}.
While the axis shape is similar to the one found in \citet{Plunk2019}, namely a racetrack oval with the points of vanishing curvature at the middle of each straight section and the surfaces resemble a {twisted strip}, there is no sudden twist near the region of maximum field strength.
This results not only in better convergence of the VMEC code at lower aspect ratios, but has a practical consequence of simplifying the coil shapes needed to produce such a configuration. As is evident from \cref{fig:elements}, not all contours of $|\mathbf B|$ close poloidally, but as we shall see below, the neoclassical transport is nevertheless very small. 

\begin{figure}
    \centering
    \includegraphics[width=.44\textwidth]{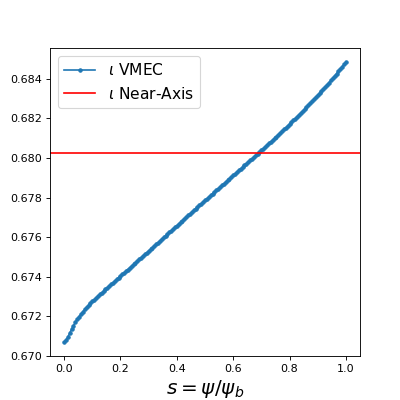}
    \includegraphics[width=.49\textwidth]{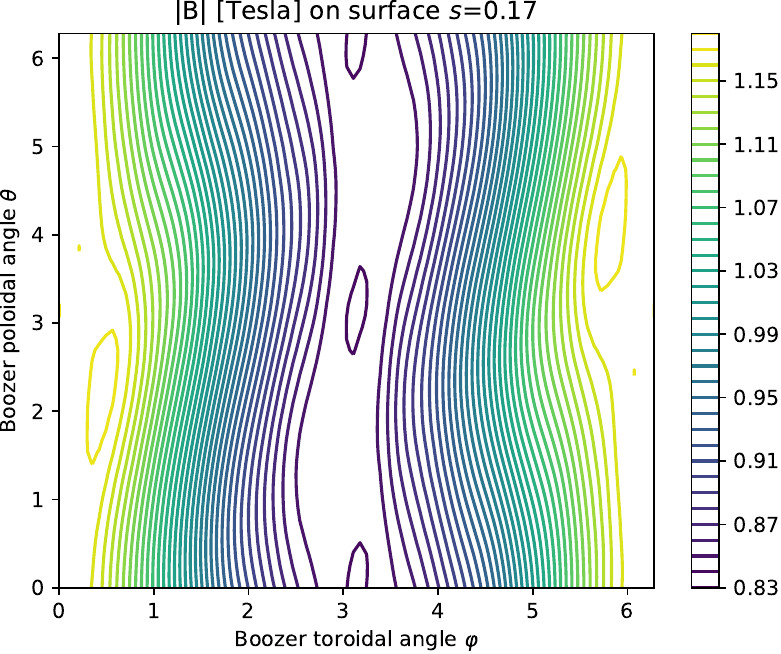}
    \includegraphics[width=.49\textwidth]{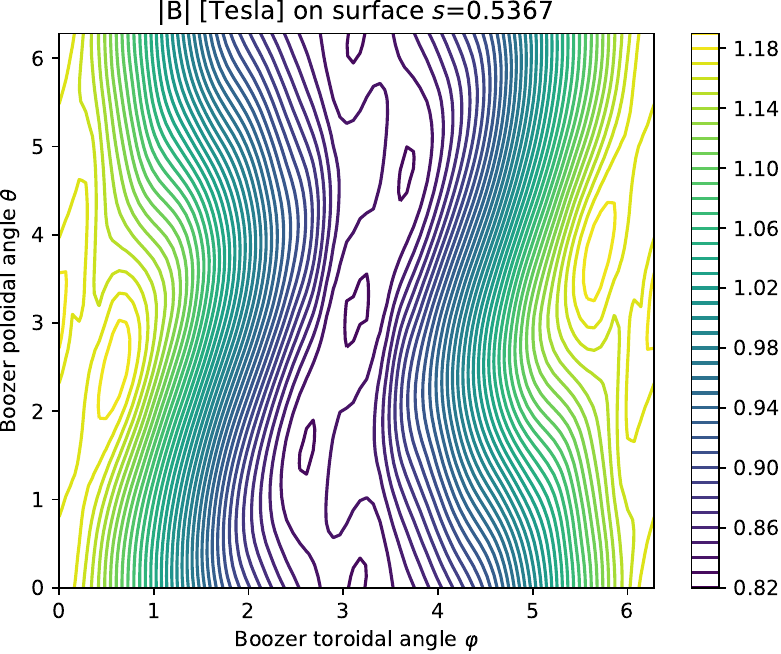}
    \includegraphics[width=.49\textwidth]{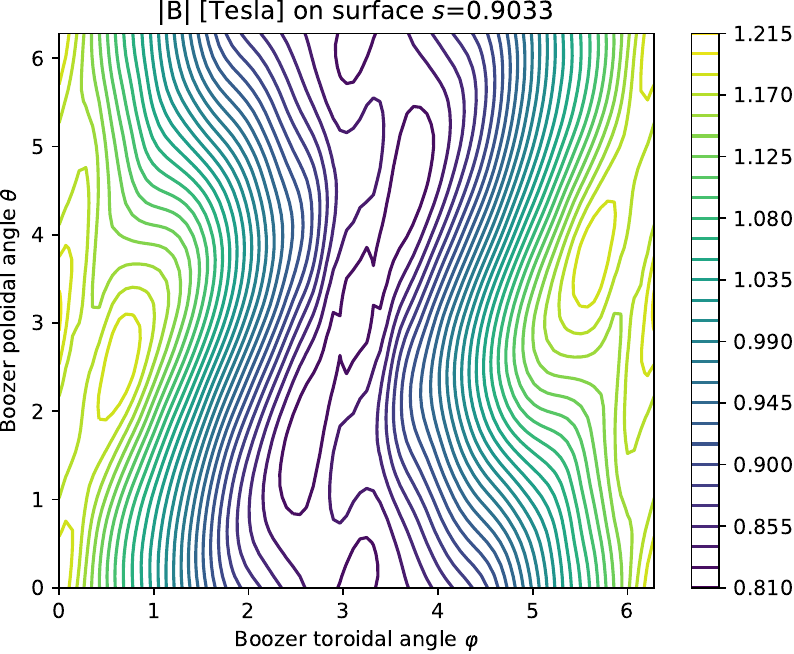}
    \caption{Top left:
    Profile of the rotational transform $\iota$ from VMEC (blue) and the $\iota$ on-axis from the near-axis expansion (red).
    Contours of constant magnetic field strength in Boozer coordinates at $s=0.17$ (top right), $s=0.5367$ (bottom left) and $s=0.9033$ (bottom right).}
    \label{fig:BOOZ_xform_contours}
\end{figure}

We then use this boundary shape as input to VMEC, a magnetic field equilibrium code.
VMEC uses an inverse moment method where the cylindrical coordinates $R$ (radial) and $Z$ (vertical) are expanded in a double Fourier series involving a poloidal angle and the cylindrical toroidal angle.
The resulting rotational transform profile is shown in Fig. \ref{fig:BOOZ_xform_contours}  (top left).
The rotational transform predicted by the near-axis expansion is $\iota=0.680$ while VMEC finds a relatively linear rotational transform with an on-axis value of $\iota=0.671$ and a value of $\iota=0.685$ at the edge.
{We note that, while the rotational transform from the near-axis expansion is fixed to the on-axis value in Fig. \ref{fig:BOOZ_xform_contours}, the inclusion of magnetic shear is possible by performing an expansion to third order as shown in \citet{Rodriguez2022}.}

Based on the VMEC result, we use the BOOZ\_XFORM code to find the magnetic field strength $|B|$ as a function of Boozer coordinates.
This allows us to draw contours of constant $|B|$ and compare with the predicted contours in \cref{fig:elements}.
The resulting properties of $|B|$ computed with BOOZ\_XFORM are shown in \cref{fig:BOOZ_xform_contours}.
In the {top right and bottom} quadrants, contours of constant $|B|$ are shown at $s=0.17,0.5367$ and $0.9033$, where $s=\psi/\psi_b$ with $\psi_b$ the toroidal magnetic flux at the boundary.

\begin{figure}
    \centering
    \includegraphics[width=.7\textwidth]{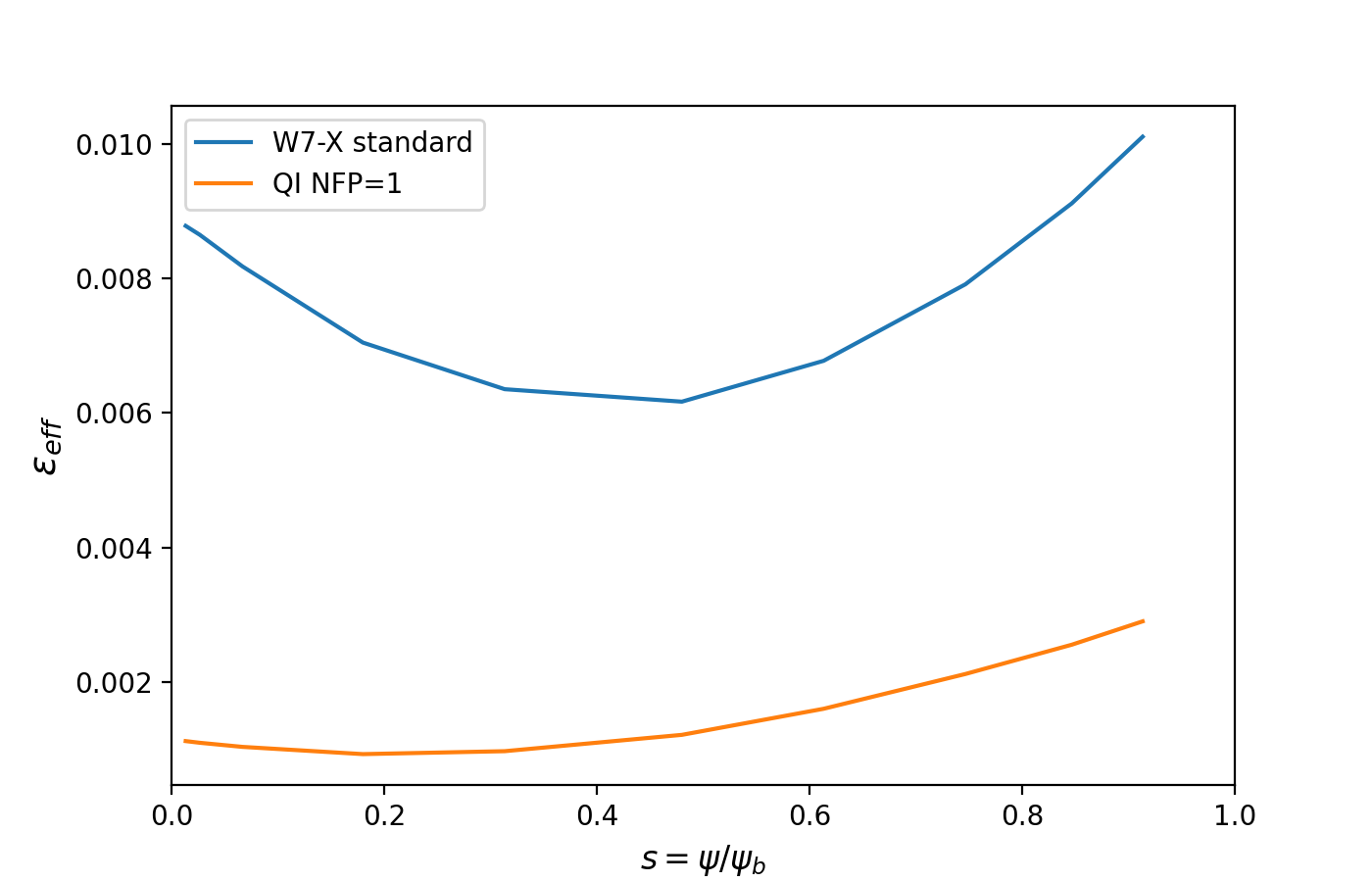}
    \caption{The magnitude $\epsilon_{\text{eff}}$ of the $1/\nu$ transport for the constructed configuration (labeled as QI NFP=1) and for the standard configuration of W7-X in fixed boundary mode.}
    \label{fig:epsiloneff}
\end{figure}

Next, we calculate the effective helical ripple, $\epsilon_{\text{eff}}$, \citep{Beidler1990,Nemov1999}, which quantifies the direct effect of the radial magnetic drift of trapped-particle orbits on neoclassical transport in the so-called 1/$\nu$ transport regime.
This parameter vanishes for perfectly omnigeneous configurations and is a function of the radial position $r^2$.
In \cref{fig:epsiloneff}, the profile of $\epsilon_{\text{eff}}$ is shown for the present configuration, calculated using the NEO code.
The $\epsilon_{\text{eff}}$ found here is substantially lower than W7-X, even for the lower aspect ratio employed here.
{For such levels of $\epsilon_{\text{eff}}$, it is expected the collisional transport to almost certainly be weaker than the turbulent transport for this configuration.}

\begin{figure}
    \centering
    \includegraphics[width=.99\textwidth]{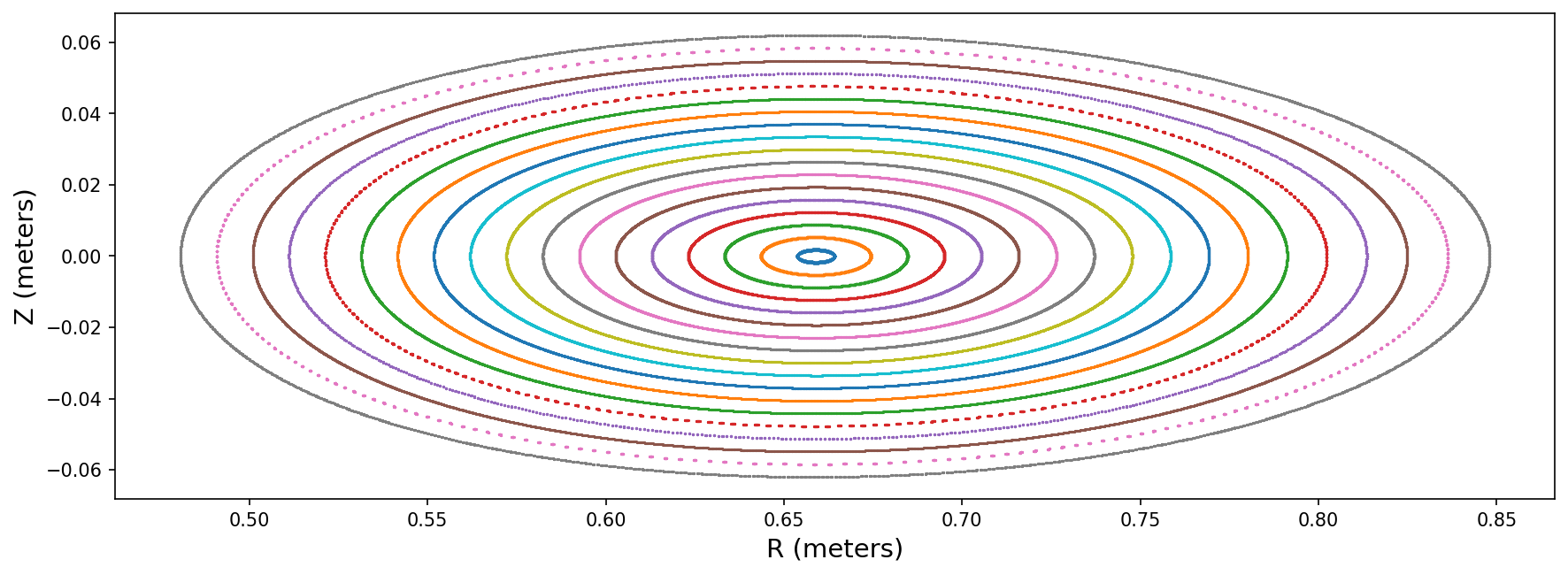}
    \caption{
    Poincaré plot computed from the SPEC solution at $\phi=0$.
    }
    \label{fig:SPEC}
\end{figure}

While VMEC can be used to analyze the approximate magnetic field inside a given boundary, it cannot be used to directly examine magnetic islands.
As discussed by \citet{Reiman2007}, low-aspect-ratio configurations are particularly prone to have magnetic islands, which are caused by undesirable radial field components at surfaces  with rational rotational transform.
As the presence of island chains and chaotic-field regions in the core significantly degrades confinement, practically useful MHD equilibria should avoid or limit this phenomenon \citep{Yamazaki1993,Neilson2010}.
In contrast to VMEC, the SPEC code is able to compute magnetic islands \citep{Hudson2012}. It does so using the formalism of multi-region relaxed MHD to divide the computational domain into a number of nested annular regions. While the magnetic field is required to be tangential to the boundary of each region, there is no requirement that magnetic surfaces exist within them. SPEC can thus be used to assess the presence of magnetic islands in the interior of such regions. For the optimized configurations found here, this procedure is particularly simple since {the} pressure vanishes and only a single region needs to be considered{, which corresponds to the vacuum problem exactly (in the numerical sense)}. 
The Poincaré plot of such a calculation using SPEC is shown in \cref{fig:SPEC}, where it is seen that no major resonances are encountered and that flux surfaces are smoothly nested from the magnetic axis to the computational boundary.

\begin{figure}
    \centering
    \includegraphics[width=.49\textwidth]{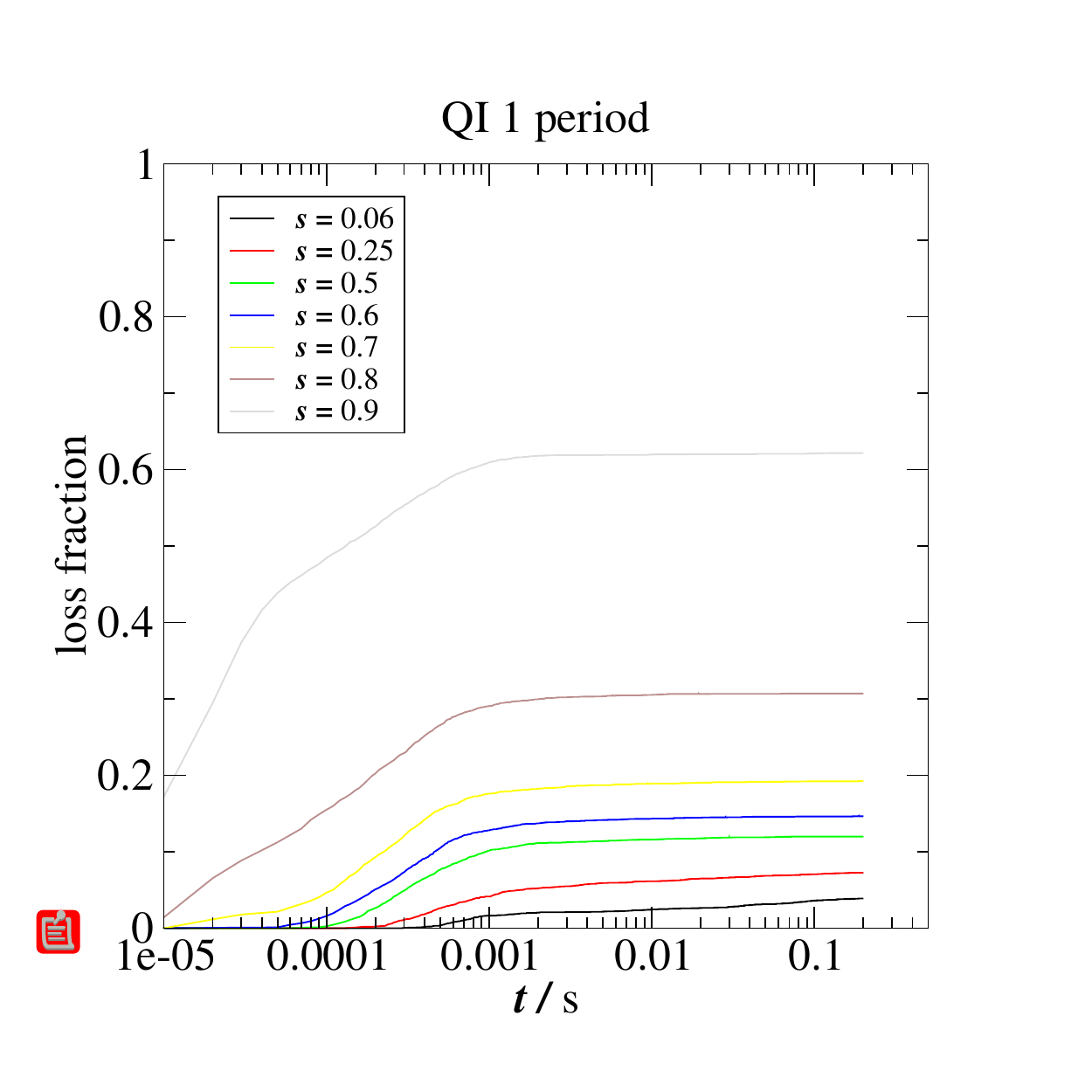}
    \includegraphics[width=.49\textwidth]{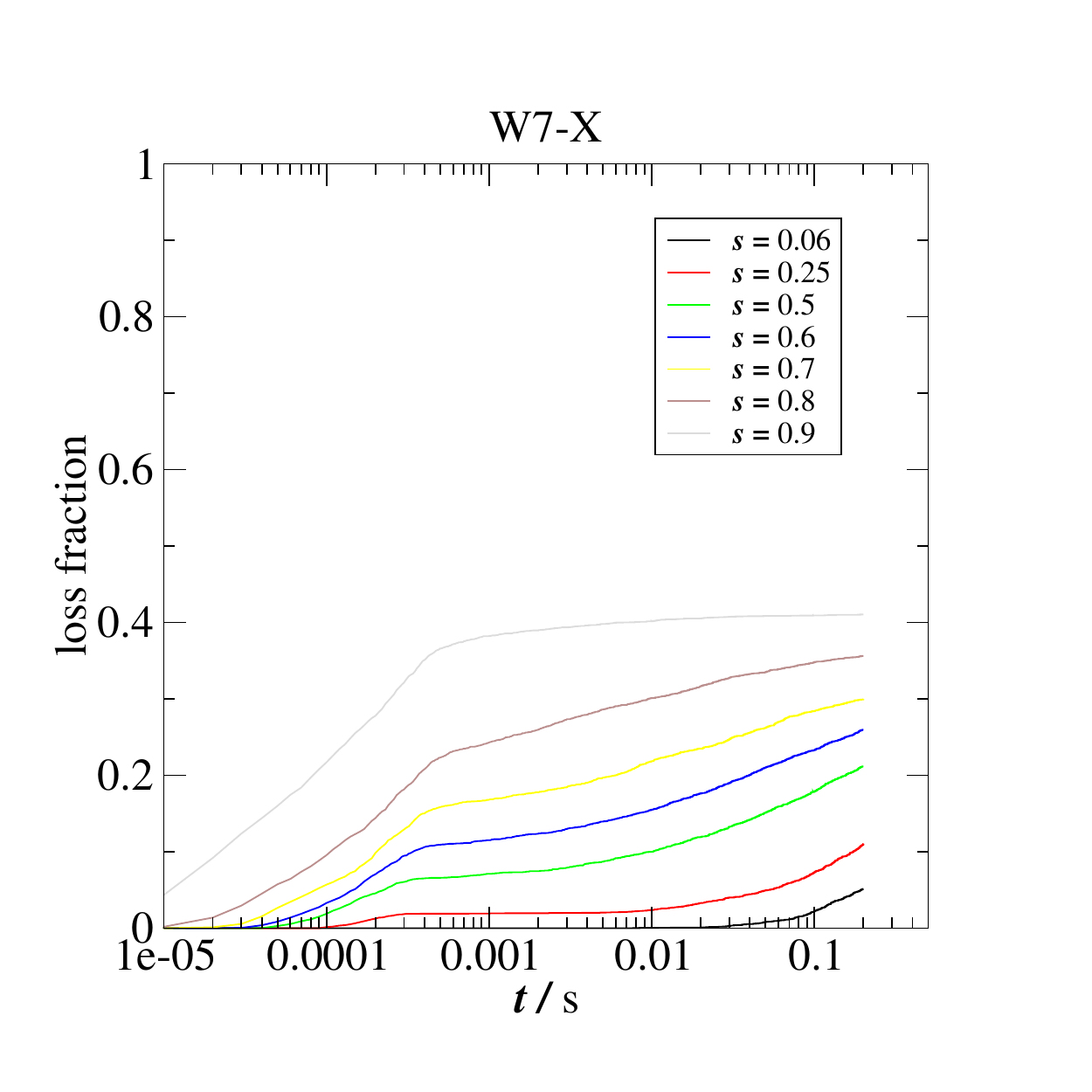}
    \caption{The fast particle loss fraction at flux surfaces between $s = 0.06$ and $s = 0.9$ for a scaled configuration to a minor radius of 1.7 m and a magnetic field $B_{0,0}=5.7 T$. Left: present 1 field period quasi-isodynamic configuration. Right: W7-X.}
    \label{fig:fast_particles}
\end{figure}

Next, we assess the confinement of 3.5 MeV alpha particles, which would be generated in a fusion reactor.
The fraction of lost particles is evaluated using the drift-orbit code SIMPLE.
For this study, we scale our magnetic configuration to a reactor size with minor radius 1.7 m and an average on-axis magnetic field strength of $B_{0,0}=5.7$ T.
Fig. \ref{fig:fast_particles} shows the loss {fraction} of fast particles following collisionless guiding-centre drift orbits. 
A total of five thousand test particles, equally distributed on each flux surface, were launched with uniformly distributed pitch angles and traced for 0.2 s, typical for the collisional slowing down time in a fusion reactor, or until they cross the s = 1 boundary surface and are considered lost.
As shown in Fig. \ref{fig:fast_particles}, the loss fraction is about 3.8\% for the particles starting at a flux surface with $s = 0.06$ and for the ones starting at a flux surface with $s = 0.25$ (approximately half radius) it is about $7.2\%$.
For comparison, Fig. \ref{fig:fast_particles} also shows the loss fraction in a W7-X configuration scaled to the same minor radius and magnetic field.
{For this study, the vacuum W7-X standard configuration is used, which corresponds to the A configuration of \citet{Geiger2015} at $\beta=0$.}
Its loss fraction is higher than the single-field-period quasi-isodynamic for all particles at $t=0.2$ s except the ones that are started at $s=0.9$.
{We note, however, that at such long time scales, collisional effects might start to play a role, making collisionless simulations less reliable. Indeed, for times $t$ smaller than 0.01 s, the single-field-period configuration provides a higher loss fraction mainly due to prompt losses. The source mechanism for such prompt losses will be the subject of future studies.}

\begin{figure}
    \centering
    \includegraphics[width=.9\textwidth]{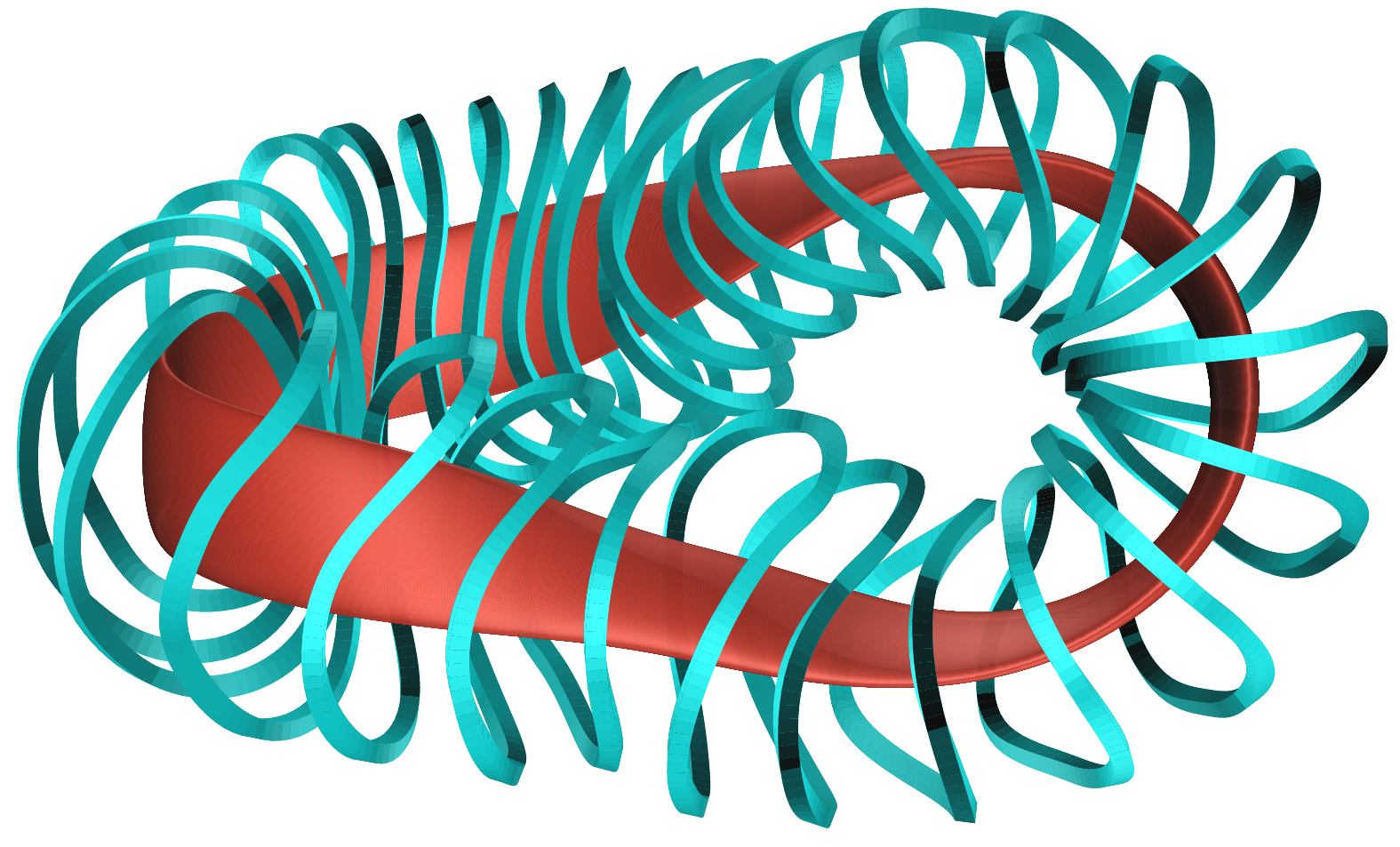}
    \caption{Coil shapes for the magnetic configuration obtained here.}
    \label{fig:coils}
\end{figure}

\begin{figure}
    \centering
    \def\svgwidth{.8\textwidth}
    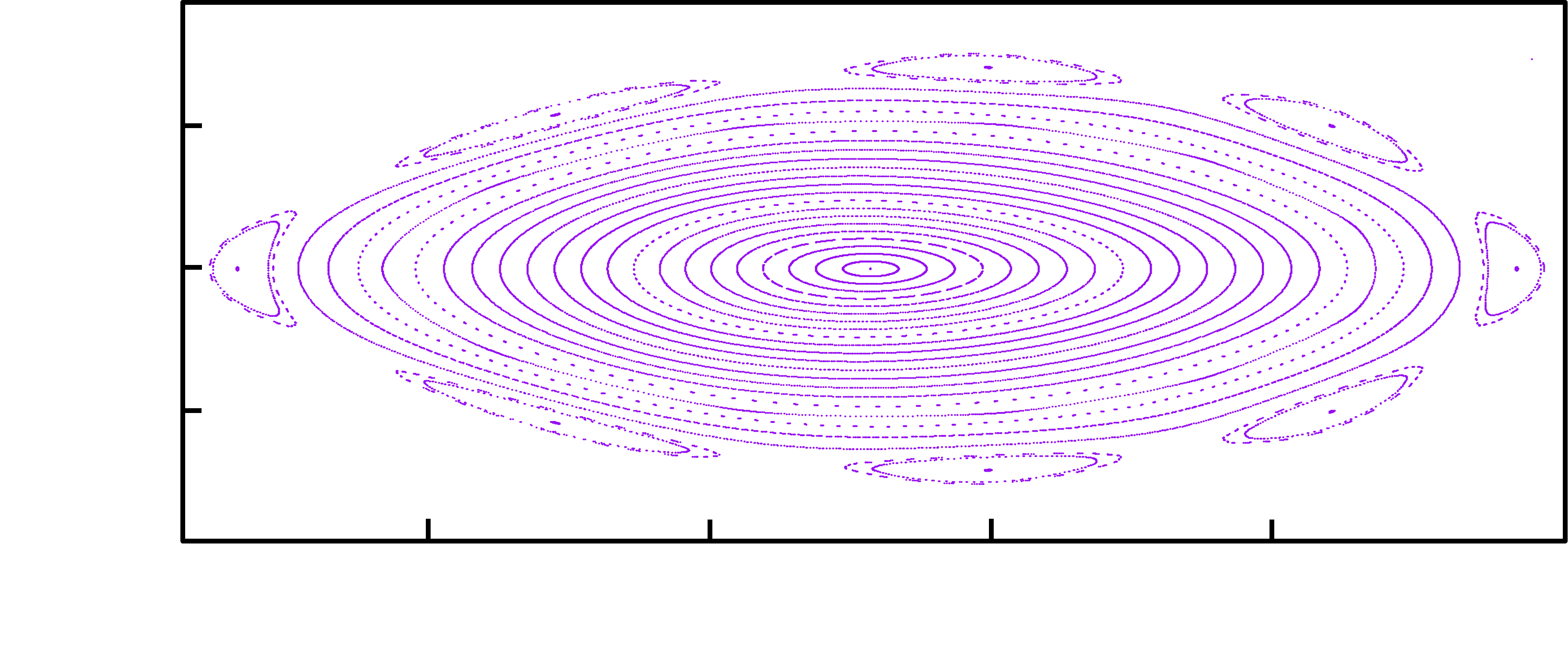
    \caption{Poincaré plot obtained with the coils shown in \cref{fig:coils} at the location $\phi=0$.}
    \label{fig:Poincare_Plot_Comparison}
\end{figure}

\begin{figure}
    \centering
    \includegraphics[width=.6\textwidth]{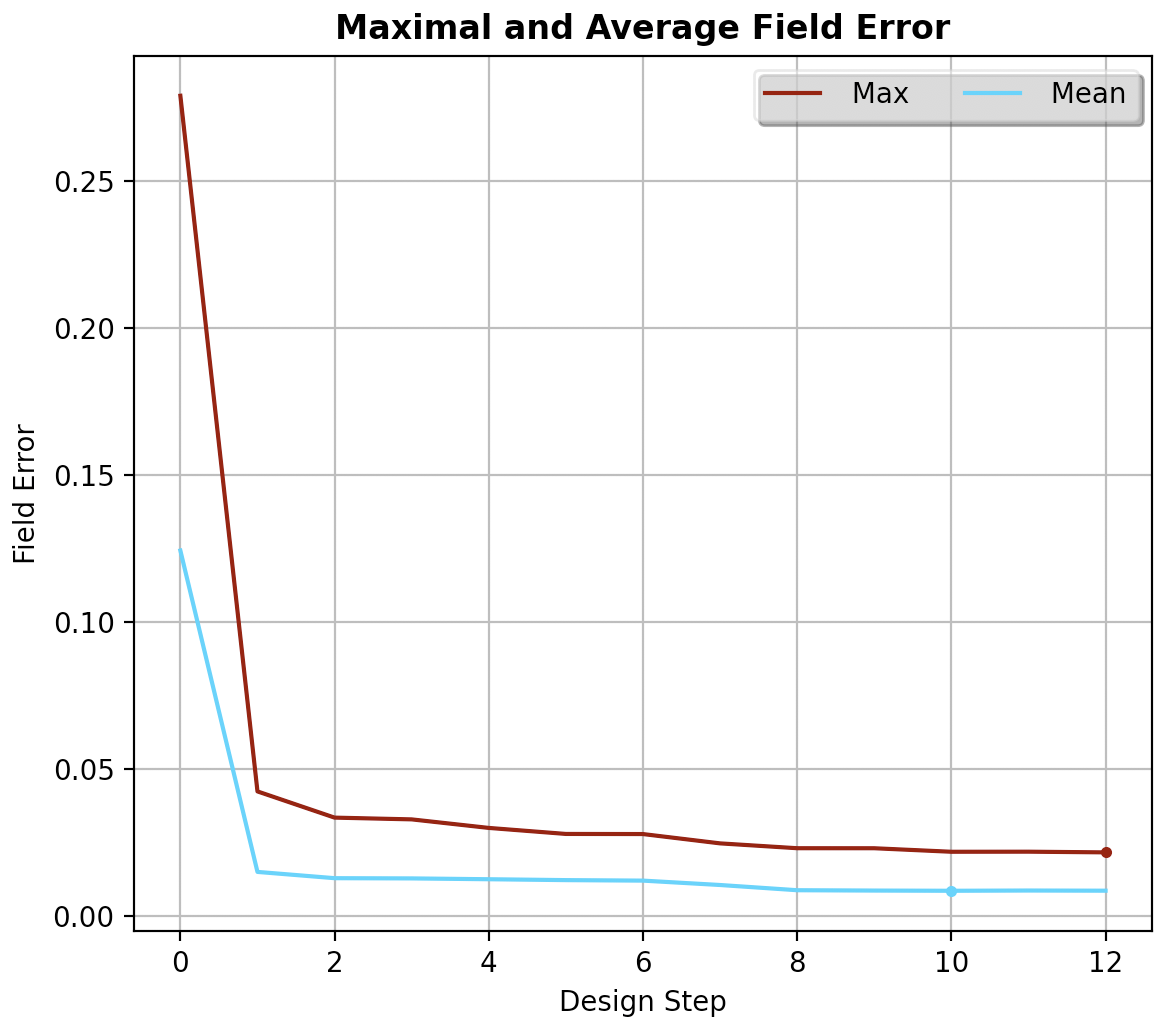}
    \caption{Relative field error {$q_{ae}$} between the magnetic field produced by the coil shapes and the VMEC boundary shape (vertical axis) as a function of the optimization step (horizontal axis).}
    \label{fig:Field_Error}
\end{figure}

Finally, a set of 30 coils that approximately reproduces the toroidal magnetic surface in \cref{fig:3d_config} was obtained.
The goal here is to show that simple coils can be found for the proposed one field period configuration.
This was done by first scanning the space of current potentials that approximate the target magnetic field. Their distance to the plasma boundary as well as the number of Fourier modes describing the analytic current can be varied. 
Then, the contours of the current potential, {found using the NESCOIL code \citep{Merkel1987},} with 4 toroidal and 3 poloidal modes that lie on a current carrying surface 30 cm away from the plasma were transformed into 15 modular coils per half-period and optimized using the ONSET code \citep{ONSET} into a constructable shape shown in \cref{fig:coils}. The coils are parametrised with 3D splines independent of any constraining surface.
The coil optimization technique used is similar to that of \citet{Lobsien2017} with the difference that a starting point was chosen that originates from a solution with higher Fourier modes. This makes the approximation of the target magnetic field better but the complexity of the starting coil configuration is prohibitive.
Therefore, a first design step of the nonlinear coil optimization focused on reducing the coil complexity as well as lowering the field error was taken. 
This procedure is also described in \citet{Lobsien2019}, which uses the same notation for the penalty values as described below.
In the optimization, the average and maximal curvature were reduced and the distance between adjacent coils was increased as well as the distance between coils and the plasma boundary.
{The minimum distance between coil centre-lines is 0.338 m}.
The final optimization step focused on properties of the vacuum magnetic field, ensuring that no low-order rational values of $\iota$ are present inside the plasma boundary and that $\iota$ increases with minor radius. 
The shapes of the magnetic surfaces defined by following field lines are highly sensitive to the iota profile due to the low shear nature of this configuration. We show in \cref{fig:Poincare_Plot_Comparison} a Poincaré plot resulting from the coil configuration in \cref{fig:coils}, which produces an island chain surrounding the last closed flux surface and could potentially be used for an island divertor. 
{We note that the appearance of an island chain outside the plasma boundary was not targeted directly.}
As seen in \cref{fig:Field_Error}, the approximation of the target magnetic field defined by the VMEC solution converges to a normalized maximum field error of 2\% and a normalized average field error below 1\% after multiple optimization steps. The normalized field error is defined locally as
\begin{eqnarray}
  \label{eq:maximumlocal}
  q_{le}(x) = \frac{|\mathbf{B}\cdot \mathbf{n}|}{|\mathbf{B}|}
\end{eqnarray}
{where $\mathbf n$ is the normal unit vector perpendicular to the plasma boundary} and the maximum local field error is $\max q_{le}$. The average field error is defined globally as
\begin{eqnarray}
  \label{eq:averageglobal}
  q_{ae}(x) =  \frac{\int_A q_{le} dA}{A},
\end{eqnarray}
where $A$ is the area of the plasma boundary.

\section{Conclusion}

In summary, a new quasi-isodynamic configuration that exhibits a number of favourable properties has been found using a first-order near-axis-expansion approach. 
It has relatively weak shaping (implying relatively simple geometry) other than strong elongation, one field period, has low neoclassical transport, and can be realized with relatively simple coils. No attempt was made to ensure favourable MHD properties, such as stability and small Shafranov shift at finite plasma pressure.

In contrast to standard optimization procedures where a plasma boundary is varied, this design was found by varying the degrees of freedom of the near-axis expansion, namely, the magnetic axis and the lowest-order magnetic field strength.
This procedure  builds on the work of \citet{Plunk2019} and \citet{Camacho2022}, using a newly developed approach to perform a controlled approximation to omnigenity.

In future work, we intend to reduce the neoclassical transport further {and improve its fast particle confinement properties} by extending the optimization procedure to second order in the inverse aspect ratio $\epsilon$.
Carrying the expansion to this order will allow us to study configurations with finite plasma pressure and to optimize for other relevant quantities such as magnetic well and the maximum-$J$-property \citep{Helander2013}.
{Indeed, this configuration is characterized by having a magnetic hill instead of a magnetic well as such measure is not available as an explicit target for the optimization at first order in the expansion.}
The particular optimization procedure outlined here can also be applied to other types of omnigenous configurations such as quasi-axisymmetric and quasi-helical symmetric ones, and will be the subject of future work.

\section{Acknowledgements}

R. J. was supported by a grant by Alexander-von-Humboldt-Stiftung, Bonn, Germany, through a research fellowship. K. C. M. and J. F. L were supported by a grant from the Simons Foundation (560651).

\bibliographystyle{jpp}
\bibliography{library}

\end{document}

%% file: Direct construction of a 1-field period quasi-isodynamic stellarator/Poincare_Plot.pdf_tex
\begingroup%
  \makeatletter%
  \providecommand\color[2][]{%
    \errmessage{(Inkscape) Color is used for the text in Inkscape, but the package 'color.sty' is not loaded}%
    \renewcommand\color[2][]{}%
  }%
  \providecommand\transparent[1]{%
    \errmessage{(Inkscape) Transparency is used (non-zero) for the text in Inkscape, but the package 'transparent.sty' is not loaded}%
    \renewcommand\transparent[1]{}%
  }%
  \providecommand\rotatebox[2]{#2}%
  \newcommand*\fsize{\dimexpr\f@size pt\relax}%
  \newcommand*\lineheight[1]{\fontsize{\fsize}{#1\fsize}\selectfont}%
  \ifx\svgwidth\undefined%
    \setlength{\unitlength}{1885.48517424bp}%
    \ifx\svgscale\undefined%
      \relax%
    \else%
      \setlength{\unitlength}{\unitlength * \real{\svgscale}}%
    \fi%
  \else%
    \setlength{\unitlength}{\svgwidth}%
  \fi%
  \global\let\svgwidth\undefined%
  \global\let\svgscale\undefined%
  \makeatother%
  \begin{picture}(1,0.42100398)%
    \lineheight{1}%
    \setlength\tabcolsep{0pt}%
    \put(0,0){\includegraphics[width=\unitlength,page=1]{Poincare_Plot.pdf}}%
    \put(0.25283563,0.04674859){\color[rgb]{0,0,0}\makebox(0,0)[lt]{\lineheight{1.25}\smash{\begin{tabular}[t]{l}$0.5$\end{tabular}}}}%
    \put(0.43132612,0.04745744){\color[rgb]{0,0,0}\makebox(0,0)[lt]{\lineheight{1.25}\smash{\begin{tabular}[t]{l}$0.6$\end{tabular}}}}%
    \put(0.61094947,0.04800226){\color[rgb]{0,0,0}\makebox(0,0)[lt]{\lineheight{1.25}\smash{\begin{tabular}[t]{l}$0.7$\end{tabular}}}}%
    \put(0.79463746,0.04800226){\color[rgb]{0,0,0}\makebox(0,0)[lt]{\lineheight{1.25}\smash{\begin{tabular}[t]{l}$0.8$\end{tabular}}}}%
    \put(0.03583033,0.15242336){\color[rgb]{0,0,0}\makebox(0,0)[lt]{\lineheight{1.25}\smash{\begin{tabular}[t]{l}$-0.1$\end{tabular}}}}%
    \put(0.05738763,0.24491131){\color[rgb]{0,0,0}\makebox(0,0)[lt]{\lineheight{1.25}\smash{\begin{tabular}[t]{l}$0.0$\end{tabular}}}}%
    \put(0.05327969,0.3377799){\color[rgb]{0,0,0}\makebox(0,0)[lt]{\lineheight{1.25}\smash{\begin{tabular}[t]{l}$0.1$\end{tabular}}}}%
    \put(-0.0012042,0.24345017){\color[rgb]{0,0,0}\makebox(0,0)[lt]{\lineheight{1.25}\smash{\begin{tabular}[t]{l}$Z$\end{tabular}}}}%
    \put(0.54181749,0.00280462){\color[rgb]{0,0,0}\makebox(0,0)[lt]{\lineheight{1.25}\smash{\begin{tabular}[t]{l}$R$\end{tabular}}}}%
  \end{picture}%
\endgroup%